\documentclass{iaus}
\usepackage{graphicx}

\title[Magnetic fields in fast rotating single giants] 
{Dynamo-generated magnetic fields in fast rotating single giants.}

\author[Konstantinova--Antova, Auriere et al.] 
{Renada Konstantinova--Antova$^1$
 Michel Auri\`ere$^2$, Klaus-Peter Schr\"oder$^3$ \and Pascal Petit$^2$}

\affiliation{$^1$Institute of Astronomy, Bulgarian Academy of Sciences,
Sofia, 72 Tsarigradsko shosse, Bulgaria \break email: renada@astro.bas.bg\\[\affilskip]
$^2$Laboratoire d'Astrophysique de Toulouse-Tarbes, Universit\'e de Toulouse, CNRS, Observatoire Midi Pyr\'en\'es,  57 
Avenue d'Azereix, 65008 Tarbes, France \break email: michel.auriere@ast.obs-mip.fr\\
$^3$Departmento de Astronomia, Universidad de Guanajuato, GTO,\break Academic Street, Camford, CF3 5QL, Mexico \break email: kps@astro.ugto.mx
}

\pubyear{2009}
\volume{259}  
\pagerange{119--126}
\date{?? and in revised form ??}
\jname{Cosmic Magnetic Fields: From Planets, to Stars and Galaxies}
\editors{K.G. Strassmeier, A.G. Kosovichev \&J.E. Beckman, eds.}
\begin{document}

\maketitle

\begin{abstract}
Red giants offer a good opportunity to study the interplay 
of magnetic fields and stellar evolution. Using the spectro-polarimeter 
NARVAL of the Telescope Bernard Lyot (TBL), Pic du Midi, France and the 
LSD technique we began a survey of magnetic fields in single G-K-M 
giants. Early results include 6 MF-detections with fast rotating giants, and
for the first time a magnetic field was detected directly in an evolved 
M-giant: EK Boo. Our results could be explained in the terms of $\alpha$--$\omega$ dynamo operating in these giants.

\end{abstract}

\firstsection 
\section{Introduction}

Magnetic fields (MF) in single evolved stars are still poorly studied. Most of 
the G--K--M giants presently been known as active are fast rotators. 
(Fekel\&Balachandran, 1993; Huensch et al., 2004). Angular momentum dredge-up
has been suggested to provide the fast rotation, driven by the convective zone
reaching near the stellar core (Simon\&Drake, 1989). In this way, MFs could be
generated by a new dynamo.
 
\section{Observations and data processing}\label{sec:obsdataproc}

The new generation spectro-polarimeters like NARVAL at TBL, Pic du Midi, 
France (Auri\`ere, 2003) and the LSD technique  (Donati et al., 1997) are 
very suitable for precision detections of MFs in giants 
(Konstantinova-Antova et al., 2008; Auri\`ere et al., 2008). We observed 7 fast 
rotating giants with NARVAL in the period of November 2006 to April 2008. 
A precise computation of the longitudinal MF $B_l$ using the First 
Moment Method (Donati et al., 1997; Rees\&Semel, 1979) was carried out, and 
the time-behavior of the activity indicator CaII K (S index) was studied.

\section{First results}\label{sec:firstresults}

\begin{figure}
 \includegraphics[height=12cm, width=4.5cm, angle=270]{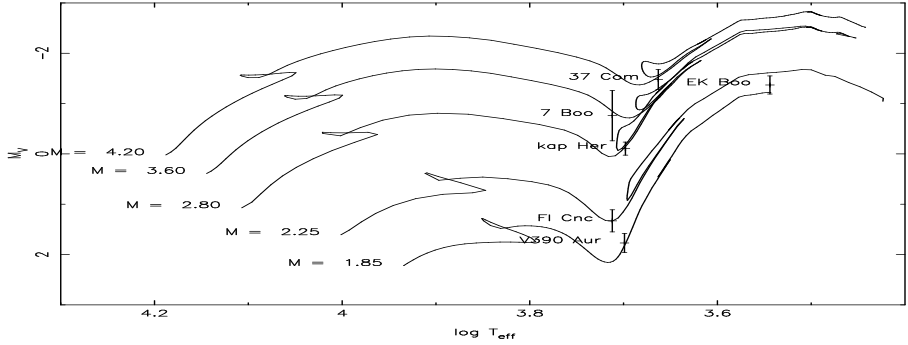}
  \caption{Situation of the fast rotating giants on the HR diagram.}

\label{fig:hr}
\end{figure}

MF structures, indicative of a dynamo, were detected for 6 of the 7 giants 
(exception: HD233517), see (incl. S index and $B_l$) Table~1. 
We determined the evolutionary status and masses of these 6 giants, using 
Hipparcos parallaxes, $T_{\rm eff}$ from the Wright catalogue (2003), 
and matching evolutionary tracks from Schr\"oder et al. (1997), 
see Figure~\ref{fig:hr}. 

\begin{table}\def~{\hphantom{0}}
  \begin{center}
  \caption{Data for the studied giants.}
  \label{tab:mf}
  \begin{tabular}{lccccccc}\hline
      $Star$  & Sp class & M/$M_\odot$ &  Vsin{i} & $L_x$ & CaII K & $B_l$ & error $B_l$ \\
              &          &        &  km/s    & $10^{30}$ erg/s & S index & Gauss & Gauss \\ \hline
       V390 Aur& G8III & 1.85 & 29 & 5.04 & 0.64 & -5 -- -15& 4.4 \\
       FI Cnc  & G8III & 2.25 & 17 & 26   & 1.04 -- 1.35 & -16.48 -- +3.45 & 1.87 \\
       37 Com  & G9III-II&4.2 & 11 & 5.20 & 0.34 & +5.62 & 0.63 \\
       7 Boo   & G5III & 3.6  & 14.5&3.72 & 0.24 & +1.83 & 0.84 \\
       $\kappa$HerA&G8III& 2.8 & 9.4 & 2.98& 0.29& -3.94 & 0.67\\
       HD233517& K2III & 1.5? & 15 &      &      & no detect.&  \\
       EK Boo  & M6III & 1.9  & 11 & 14.12& 0.21--0.26& -3.19 -- -6.76& 0.62\\\hline
  \end{tabular}
 \end{center}
\end{table}

All 6 giants with MF-detection have masses $\ge$1.5 $M_\odot$ and their 
convective zones are currently deepening while the stars are evolving. Their
evolutionary stages reach from the Hertzsprung gap to the AGB. 
While the particular reasons for the fast rotation could be different, 
depending on mass and evolutionary history, an $\alpha$--$\omega$ dynamo 
presents a likely reason for the detected magnetic activity.

\begin{acknowledgments}
We thank the TBL team for providing the observations, a substantial part 
was supported by the OPTICON programme. The Hipparcos database was used.
\end{acknowledgments}

\end{document}